\documentclass[aps,prl,superscriptaddress,twocolumn]{revtex4-2}
\usepackage{graphicx}
\usepackage{color}
\usepackage{bm}
\usepackage{braket}
\usepackage{hyperref}
\usepackage{lipsum}
\usepackage{float}

\begin{document}

\title{High-Energy Molecular-Frame Photoelectron Angular Distributions:\\ A Molecular Bond-Length Ruler}

\author{I.~Vela-Per\'ez}
\affiliation{Institut f\"ur Kernphysik, Goethe-Universit\"at, Max-von-Laue-Strasse 1, 60438 Frankfurt am Main, Germany}

\author{F.~Ota}
\affiliation{Department of Physics, University of Toyama, Gofuku 3190, Toyama 930-8555, Japan}

\author{A.~Mhamdi}
\affiliation{Institut f\"ur Physik und CINSaT, Universit\"at Kassel, Heinrich-Plett-Strasse 40, 34132 Kassel, Germany}

\author{Y.~Tamura}
\affiliation{Department of Physics, University of Toyama, Gofuku 3190, Toyama 930-8555, Japan}

\author{J.~Rist}
\affiliation{Institut f\"ur Kernphysik, Goethe-Universit\"at, Max-von-Laue-Strasse 1, 60438 Frankfurt am Main, Germany}

\author{N.~Melzer}
\affiliation{Institut f\"ur Kernphysik, Goethe-Universit\"at, Max-von-Laue-Strasse 1, 60438 Frankfurt am Main, Germany}

\author{S.~Uerken}
\affiliation{Institut f\"ur Kernphysik, Goethe-Universit\"at, Max-von-Laue-Strasse 1, 60438 Frankfurt am Main, Germany}

\author{G.~Nalin}
\affiliation{Institut f\"ur Kernphysik, Goethe-Universit\"at, Max-von-Laue-Strasse 1, 60438 Frankfurt am Main, Germany}

\author{N.~Anders}
\affiliation{Institut f\"ur Kernphysik, Goethe-Universit\"at, Max-von-Laue-Strasse 1, 60438 Frankfurt am Main, Germany}

\author{D.~You}
\affiliation{Institute of multidisciplinary research for advanced materials, Tohoku University, Sendai 980-8577, Japan}

\author{M.~Kircher}
\affiliation{Institut f\"ur Kernphysik, Goethe-Universit\"at, Max-von-Laue-Strasse 1, 60438 Frankfurt am Main, Germany}

\author{C.~Janke}
\affiliation{Institut f\"ur Kernphysik, Goethe-Universit\"at, Max-von-Laue-Strasse 1, 60438 Frankfurt am Main, Germany}

\author{M.~Waitz}
\affiliation{Institut f\"ur Kernphysik, Goethe-Universit\"at, Max-von-Laue-Strasse 1, 60438 Frankfurt am Main, Germany}

\author{F.~Trinter}
\affiliation{Deutsches Elektronen-Synchrotron (DESY), Notkestrasse 85, 22607 Hamburg, Germany}
\affiliation{Molecular Physics, Fritz-Haber-Institut der Max-Planck-Gesellschaft, Faradayweg 4-6, 14195 Berlin, Germany}

\author{R.~Guillemin}
\affiliation{Sorbonne Universit\'e CNRS, Laboratoire de Chimie Physique-Matiere et Rayonnement, LCPMR, F-75005, Paris, France}

\author{M.~N.~Piancastelli}
\affiliation{Sorbonne Universit\'e CNRS, Laboratoire de Chimie Physique-Matiere et Rayonnement, LCPMR, F-75005, Paris, France}
\affiliation{Department of Physics and Astronomy, Uppsala University, SE-751 20 Uppsala, Sweden}

\author{M.~Simon}
\affiliation{Sorbonne Universit\'e CNRS, Laboratoire de Chimie Physique-Matiere et Rayonnement, LCPMR, F-75005, Paris, France}

\author{V.~T.~ Davis}
\affiliation{Department of Physics, University of Nevada, Reno, Nevada 89557, USA}

\author{J.~B.~ Williams}
\affiliation{Department of Physics, University of Nevada, Reno, Nevada 89557, USA}

\author{R.~D\"orner}
\affiliation{Institut f\"ur Kernphysik, Goethe-Universit\"at, Max-von-Laue-Strasse 1, 60438 Frankfurt am Main, Germany}

\author{K.~Hatada}
\affiliation{Department of Physics, University of Toyama, Gofuku 3190, Toyama 930-8555, Japan}

\author{K.~Yamazaki}
\affiliation{RIKEN Center for Advanced Photonics, RIKEN, 2-1 Hirosawa, Wako, Saitama, 351-0198, Japan}

\author{K.~Fehre}
\affiliation{Institut f\"ur Kernphysik, Goethe-Universit\"at, Max-von-Laue-Strasse 1, 60438 Frankfurt am Main, Germany}

\author{Ph.~V.~Demekhin}
\affiliation{Institut f\"ur Physik und CINSaT, Universit\"at Kassel, Heinrich-Plett-Strasse 40, 34132 Kassel, Germany}

\author{K.~Ueda}
\affiliation{Institute of multidisciplinary research for advanced materials, Tohoku University, Sendai 980-8577, Japan}
\affiliation{Department of Chemistry, Tohoku University, 6-3 Aramaki Aza-Aoba, Aoba-ku, Sendai 980-8578, Japan}

\author{M.~S.~Sch\"offler}
\affiliation{Institut f\"ur Kernphysik, Goethe-Universit\"at, Max-von-Laue-Strasse 1, 60438 Frankfurt am Main, Germany}

\author{T.~Jahnke}
\affiliation{European XFEL, Holzkoppel 4, 22869 Schenefeld, Germany}

\date{\today}

\begin{abstract}
We present an experimental and theoretical study of core-level ionization of small hetero- and homo-nuclear molecules employing circularly polarized light and address molecular-frame photoelectron angular distributions in the light's polarization plane (CP-MFPADs). We find that the main forward-scattering peaks of CP-MFPADs are slightly tilted with respect to the molecular axis. We show that this tilt angle can be directly connected to the molecular bond length by a simple, universal formula. The extraction of the bond length becomes more accurate as the photoelectron energy is increased. We apply the derived formula to several examples of CP-MFPADs of C 1s and O 1s photoelectrons of CO, which have been measured experimentally or obtained by means of {\it ab initio} modeling. The photoelectron kinetic energies range from 70 to 1000~eV and the extracted bond lengths agree well with the known bond length of the CO molecule in its ground state. In addition, we discuss the influence of the back-scattering contribution that is superimposed over the analyzed forward-scattering peak in case of homo-nuclear diatomic molecules as N$_2$.
\end{abstract}

\maketitle

Recent developments of X-ray free-electron lasers (XFELs) delivering extremely short X-ray pulses of a few femtoseconds~\cite{Emma2010} and of a  mega-electron-volt pulsed electron beam~\cite{Weathersby2015} at SLAC have paved new pathways to image structural changes of molecules in chemical reactions, for example, employing time-resolved X-ray diffraction~\cite{Minitti2015} and ultrafast electron diffraction~\cite{Wolf2019}. While catching the motion of the individual atoms, including hydrogen atoms, of single molecules during chemical reactions is of fundamental interest, it still remains a challenge.

A couple of decades ago, it was pointed out that core-level photoelectron angular distributions include molecular structure information, which can be accessed if the molecule is either fixed in space or if the photoelectron angular distribution in the molecular frame (MFPAD) is retrieved from a coincidence measurement~\cite{Landers2001}. Accordingly, measuring the MFPAD of an isolated molecule is the gas-phase analog to photoelectron diffraction imaging (PED) routinely used for studying surface structure~\cite{Woodruff2008}. Following the spirit of PED, a couple of groups~\cite{Krasniqi2010, Kazama2013} proposed time-resolved MFPAD measurements using an XFEL light pulse as an ionizing source, as one of the future routes to image the motion of individual atoms in a molecule. Several attempts towards time-resolved MFPAD measurements with XFELs were reported, in which the sample molecules were aligned by employing an impulsive or adiabatic alignment method using optical laser fields~\cite{Rouzee2013, Boll2013, Nakajima2015, Minemoto2016}. However, no time-resolved studies of molecular structural changes  have been reported, so far, partly due to the complexity of the needed pulse sequence: A first laser pulse is needed to align the molecule, a second laser pulse triggers the photoreaction, and a final XFEL pulse ejects the photoelectron at a particular site in the molecule in order to perform PED.

The first high-repetition-rate XFEL, the European XFEL~\cite{Decking2020}, opened the door to coincidence experiments using COLTRIMS (Cold Target Recoil Ion Momentum Spectroscopy) reaction microscopes (REMI)~\cite{Ullrich2003} for MFPAD measurements. In a COLTRIMS measurement, the spatial orientation of a molecule can be deduced after triggering a fragmentation of the molecule and detecting the momenta of the fragment ions in coincidence. As the momentum of the ejected photoelectron is recorded in coincidence with the fragment ions, the photoelectron's emission direction with respect to the molecule can be determined.  Very recently, Kastirke {\it et al.} have reported a first successful implementation of this technique at the soft X-ray beamline of the European XFEL~\cite{Kastirke2020,Kastirke2020_2}.

To establish MFPAD measurements as a routinely usable tool for molecular structure imaging, however, an important ingredient is still missing, which is a method to extract the molecular structure directly from the measured MFPAD without the need of a full theoretical modeling of the photoemission process. So far, concepts proposed for achieving this goal focus on polarization-averaged MFPADs (see, e.g.,~\cite{Ota2021a} and references therein), and rely on two prerequisites: Firstly, at sufficiently high photoelectron kinetic energies for which a single-scattering approximation is valid (say, above 70~eV), forward-scattering peaks  observable in polarization-averaged MFPADs coincide with the relative location (from the point of view of the emitter atom) of neighboring atoms~\cite{Williams2012,Plesiat2013}. Secondly, the molecular-frame interference pattern caused by the direct and scattered photoelectron waves can be correlated to the distance between the emitter and the scatterers~\cite{Fukuzawa2019, Ota2020paper2}. In addition, in case of homo-nuclear molecules, bond lengths information can be obtained exploiting the analogy between Young's double slit principle and MFPADs (see e.g. \cite{Kushawaha15201}).

In the present work, we demonstrate a different approach building on the idea of stereo-atomscope diffraction spectroscopy known in the field of surface-structure determination, in which circularly polarized light is employed as an ionizing source for PED~\cite{Daimon2001,Matsui2010}. Namely, employing the hetero-nuclear diatomic molecule CO as a prototype system, we show that high-energy MFPADs of fixed-in-space molecules provide direct access to structural features such as the molecule's bond length, if circularly polarized light is used for the ionization and the MFPAD is confined to the polarization plane. Such MFPADs are hereafter referred to as CP-MFPADs.

The remaining part of the manuscript is structured as follows: First, we present the experimental C 1s and O 1s CP-MFPADs of CO. We find that the high-energy CP-MFPAD exhibits petal-like features (i.e., a flower shape) with a strong forward-scattering peak in the direction from the emitter to the scatterer. The latter peak is slightly tilted upward or downward, depending on the handedness of the circularly polarized photons. We then show that our {\it ab initio} calculations reproduce very well the experimental CP-MFPADs. Finally, we demonstrate that the features of high-energy CP-MFPADs described above can be well captured by a very simple analytic expression, and that the bond length can be directly read off from the tilt angle of the forward-scattering peak without any need for \textit{ab initio} modeling. For comparison and further insight, we present also the results for the homo-nuclear diatomic molecule N$_2$.

The experiment was performed at Synchrotron SOLEIL (Saint-Aubin, France) using the well-established COLTRIMS technique~\cite{Ullrich2003} during the 8-bunch mode (pulsed operation) of the synchrotron. A supersonic expansion of CO or N$_2$ gas was skimmed to form a molecular beam that crossed the synchrotron radiation provided by the variable-polarization undulator-based beamline SEXTANTS. Electrons and ions generated by photoionization and subsequent Auger decay were accelerated (in opposite directions) by a static electric field (38.3 V/cm), onto two position- and time-sensitive microchannel plate detectors (MCPs) with multi-hit-capable delay-line position readout \cite{jagutzki02nim}. From the position of impact on the detectors, the known distance between the ionization region and the MCPs, and the time-of-flight, the particle trajectories inside the spectrometer can be deduced, and from these, the initial momentum vector of each particle can be determined. The electron spectrometer of the COLTRIMS analyzer was specifically designed to allow for a  measurement of high-energy photoelectrons (up to 1~keV) by limiting the acceptance angle for electron detection. 
For each ionization event, a core-level photoelectron was detected in coincidence with the two fragment ions resulting from the Coulomb explosion of the molecule after Auger decay.

\begin{figure}
\includegraphics[width=1.0\linewidth]{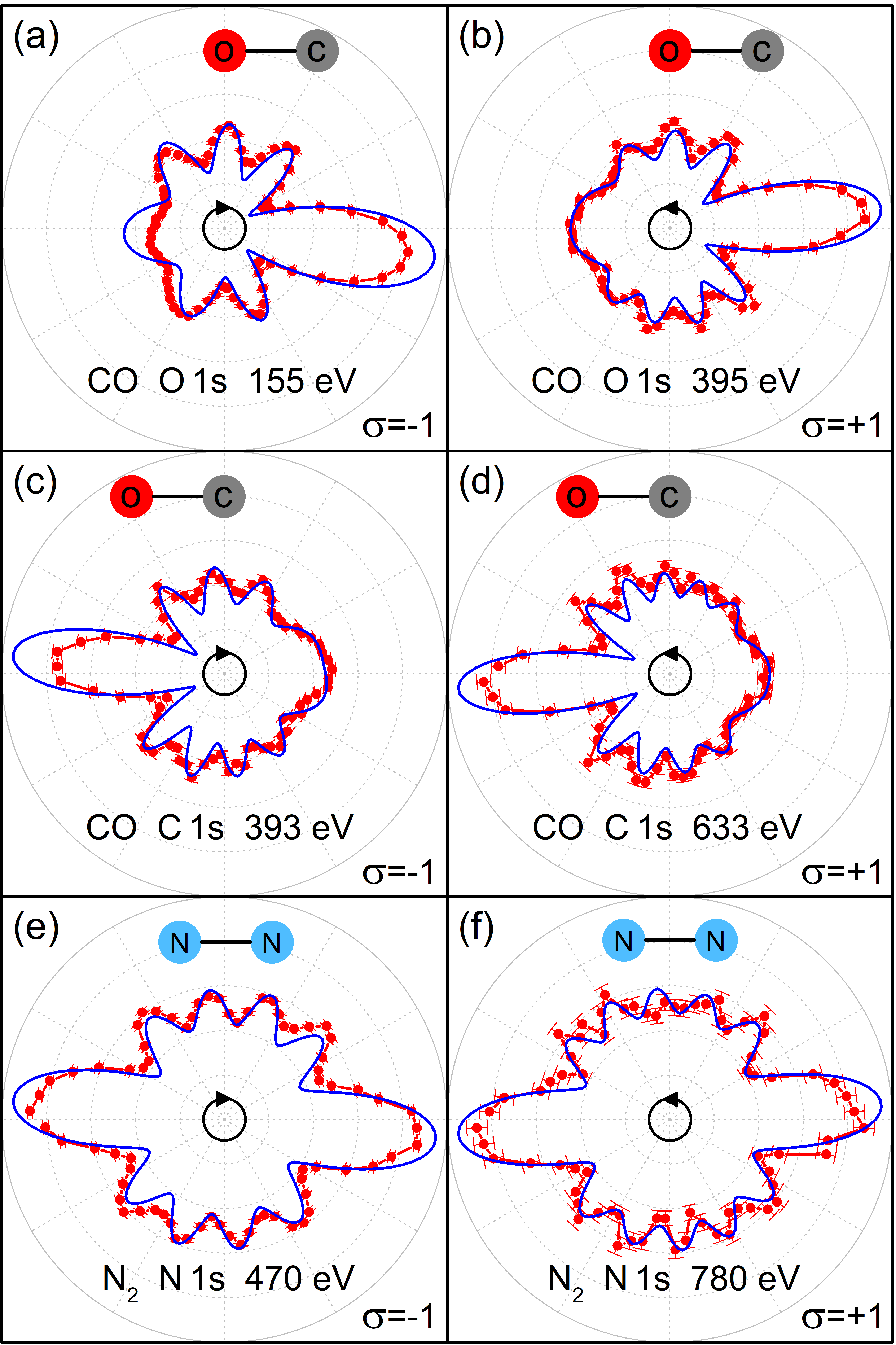}
\caption{
Comparison between the measured (red dots with error bars) and calculated (blue solid lines) CP-MFPADs for different photoelectron kinetic energies.
(a) CO O 1s at 155~eV,
(b) CO O 1s at 395~eV,
(c) CO C 1s at 393~eV,
(d) CO C 1s at 633~eV,
(e) N$_2$ N 1s at 470~eV, and
(f) N$_2$ N 1s at 780~eV.
}
\label{fig:exp_calc}
\end{figure}

For our {\it ab initio} modeling, we employed the following approach: The angular distributions of the C 1s and O 1s photoelectrons of the CO molecule and of the N 1s photoelectrons of the N$_2$ molecule were computed by using the single-center (SC) method and code \cite{code}, which provides an accurate description of the partial photoelectron continuum waves in molecules. The calculations were carried out at the equilibrium internuclear geometries of the ground electronic states of CO and N$_2$ in the frozen-core Hartree-Fock approximation.  The SC expansions of the occupied orbitals were restricted by partial harmonics with $\ell_c \leq 99$ and for  photoelectrons with  $\ell_\varepsilon\leq 49$.

We performed measurements using right circularly and left circularly polarized light at photon energies of $h\nu=690$~eV and $h\nu=930$~eV for the studies on CO, and $h\nu=880$~eV and $h\nu=1190$~eV for the investigation on N$_2$. The resulting photoelectron kinetic energies are 393~eV and 633~eV for C 1s photoemission of CO, 155~eV and 395~eV for O 1s photoemission of CO, and 470~eV and 780~eV for N 1s photoemission of N$_2$. Figure~\ref{fig:exp_calc} depicts the respective experimental CP-MFPADs and those obtained from our {\it ab initio} calculations. Our modeling nicely reproduces all features observed in the experiment. Looking at the CO results in Figs.~\ref{fig:exp_calc} (a) to (d), we find that the number of petals (resembling a flower shape), which appear as a result of interference between the direct and scattered photoelectron waves, increases as photoelectron energy increases. The same behavior has been seen in studies on polarization-averaged MFPADs~\cite{Plesiat2013}.

The most striking feature of all the CP-MFPADs is a very strong forward-scattering peak which  appears in the direction of the scatterer atom that is neighboring the emitter atom: The forward scattering is not located exactly along the molecular axis but occurs slightly tilted from it in the direction of rotation of the electric field vector. This upward or downward tilt thus flips if the handedness of the ionizing light is changed from the right ($\sigma=-1$) to the left ($\sigma=1$). This tilt is a characteristic feature of the CP-MFPADs, as also of PED on surfaces~\cite{Daimon2001,Matsui2010}. Furthermore, close inspections reveal that the tilt angle decreases  with the increase of the photoelectron energy.  The results of N$_2$ in Figs.~\ref{fig:exp_calc} (e) and (f) depict these features as well.

\begin{figure}
\includegraphics[width=1.0\linewidth]{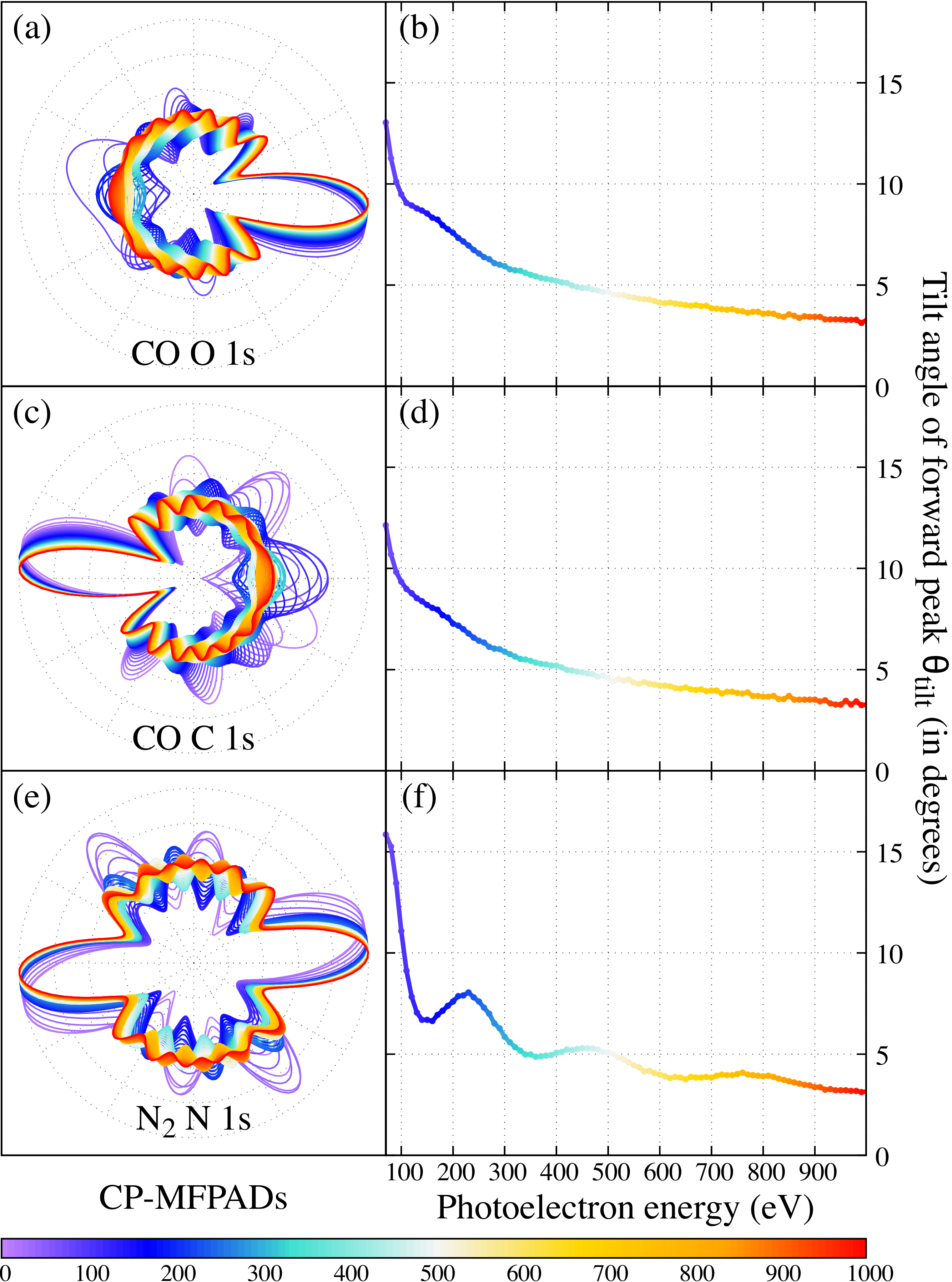}
\caption{CP-MFPADs resulting from our {\it ab initio} calculations for (a) CO O 1{s}, (c) CO C 1{s}, and (e) N$_{2}$ N 1{s} as a function of photoelectron kinetic energy in a range from 70 to 1000~eV. The tilt angles of the forward peak, $\theta_\mathrm{tilt}$, of the CP-MFPADs are displayed as functions of the photoelectron kinetic energy for (b) CO O 1{s}, (d) CO C 1{s}, and (f) N$_{2}$ N 1{s}, respectively. The color encodes the photoelectron energy in all panels.
}
\label{fig:tilt}
\end{figure}

In order to gather more detailed information on the findings described above, we performed further {\it ab initio} calculations over a wide range of  photoelectron energies from 70 eV up to 1000 eV. The results are summarized in Fig.~\ref{fig:tilt}. Figures~\ref{fig:tilt}(a), \ref{fig:tilt}(c), and \ref{fig:tilt}(e) illustrate that, indeed, the number of  petals in the flower shape of the CP-MFPAD increases while the tilt angle of the forward-scattering peak decreases with the increase of the photoelectron kinetic energy.  Figures~\ref{fig:tilt}(b), \ref{fig:tilt}(d), and \ref{fig:tilt}(f) depict the extracted tilt angle $\theta_{\rm tilt}$ as a function of photoelectron energy. For the CO molecule (Figs.~\ref{fig:tilt}(b) and \ref{fig:tilt}(d)), we observe monotonic decreases of the tilt angle with the increase of the photoelectron energy. The results for N$_2$ depicted in Fig.~\ref{fig:exp_calc}(f) show that the decrease in the tilt angle is additionally modulated by a damped oscillation.

Aiming at elucidating these characteristic behaviors of the measured and calculated CP-MFPADs, we derive analytic expressions for describing the CP-MFPAD of hetero-nuclear diatomic molecules AB in the high-energy regime. We choose atom A as the photo-absorbing atom that emits a photoelectron. We define the bond length vector $\mathbf{R}=\mathbf{r}_\mathrm{B}-\mathbf{r}_\mathrm{A}$, the photoelectron  wave vector $\hat{\mathbf{k}}$, and angle $\theta$ between them. We assume that CP-MFPADs are measured in the polarization plane of the circular light and employ the electric-dipole, single-channel, and single-scattering approximations using a site \textit{T}-matrix expansion~\cite{Faulkner1980,hatada2010}. In this case, the  CP-MFPAD as a function of  photoelectron emission vector, ${\bf k }$, is given by the following form~\cite{Ota2021a,Ota2020paper2}:
\begin{eqnarray}
&I^{A}\left(\bf{k},\hat{\varepsilon}_{\sigma}\right)
\propto
\left| t^{A}_{1} \right|^2 \displaystyle\left( \frac{k}{\pi} \right) \displaystyle\frac{3}{4\pi}
\Bigg\{
1 + \displaystyle\frac{\left|f(k,\theta)\right|^2}{R^2} \nonumber\\
&+
2\displaystyle\frac{\left|f(k,\theta)\right|}{R}
\cos\left[kR\left(1-\cos\theta\right)+\phi(k,\theta)-\sigma\theta\right]
\Bigg\},
\label{eqn:CP-MFPADs}
\end{eqnarray}
where we defined the scattering amplitude of atom B as $f(k,\theta) =\left|f(k,\theta)\right|e^{i\phi(k,\theta)}$
and $\phi(k,\theta)$ is a phase function. The polarization index $\sigma$=$-1$ and $+1$ indicates right- and left-handed polarization, respectively.
The first and the second terms of Eq.~(\ref{eqn:CP-MFPADs}) are the direct wave from the emitter A and the scattering wave from the scatterer B, respectively.
The third term represents the interference between the direct and single-scattered waves and depends on the bond length.

Constructive and destructive interference in the differential photoemission intensity (i.e., the fringes in the CP-MFPAD) correspond to maxima and minima of the following fringe function:
\begin{equation}
G\left(\theta\right) =
\cos
\left[
kR \left(1-\cos\theta \right) + \phi(\theta) - \sigma \theta
\right].\label{eqn:fringe}
\end{equation}
Here, we omitted the argument $k$ of $\phi$. In order to find the angles at which the maxima and minima appear in $G(\theta)$, we set $dG(\theta)/d\theta = 0$:
\begin{eqnarray}
\frac{dG(\theta)}{d\theta}
&=&
-\sin
\left[kR \left(1-\cos\theta \right) + \phi(\theta) - \sigma \theta\right]\nonumber\\
&&\times
\left[kR \sin\theta + \frac{d\phi(\theta)}{d\theta} - \sigma\right] = 0.
\label{eqn:dgdtheta}
\end{eqnarray}
This condition is satisfied either when the sine function becomes zero, i.e., $\sin\left[kR \left(1-\cos\theta \right) + \phi(\theta) - \sigma \theta\right] = 0$  or when the second term in braces vanishes, i.e., $kR \sin\theta + d\phi(\theta)/d\theta - \sigma=0$. The first condition describes the angles at which maxima and minima form the flower shape in the CP-MFPAD shown in Figs.~\ref{fig:exp_calc}(a) to \ref{fig:exp_calc}(d) and \ref{fig:tilt}(a) and \ref{fig:tilt}(b)). The second condition yields a relationship between $kR$ and the tilt angle $\theta_{\rm tilt}$. This relation confirms that the tilt angle decreases with the increase of the photoelectron momentum $k$ (and thus, of the photoelectron kinetic energy), as depicted in Fig.~\ref{fig:tilt}(b) and \ref{fig:tilt}(d) for CO.
If $|f(k,\theta)|_{\theta=0,\,\pi}\ne0$, as in this study, $d\phi(\theta)/d\theta |_{\theta=0,\,\pi}=0$ holds~\cite{Farrell1971}.
We can thus assume that $d\phi(\theta)/d\theta$ is negligible for small tilt angles from the forward and backward directions. We finally obtain the following relationship:
\begin{equation}
R_{\rm Hetero}^{\rm tilt}
= \frac{\sigma}{k \sin(\theta_{\rm tilt})}=\frac{1}{k \sin(|\theta_{\rm tilt}|)}\;.
\label{eqn:tilt_hetero}
\end{equation}

\begin{figure}
\includegraphics[width=1.0\linewidth]{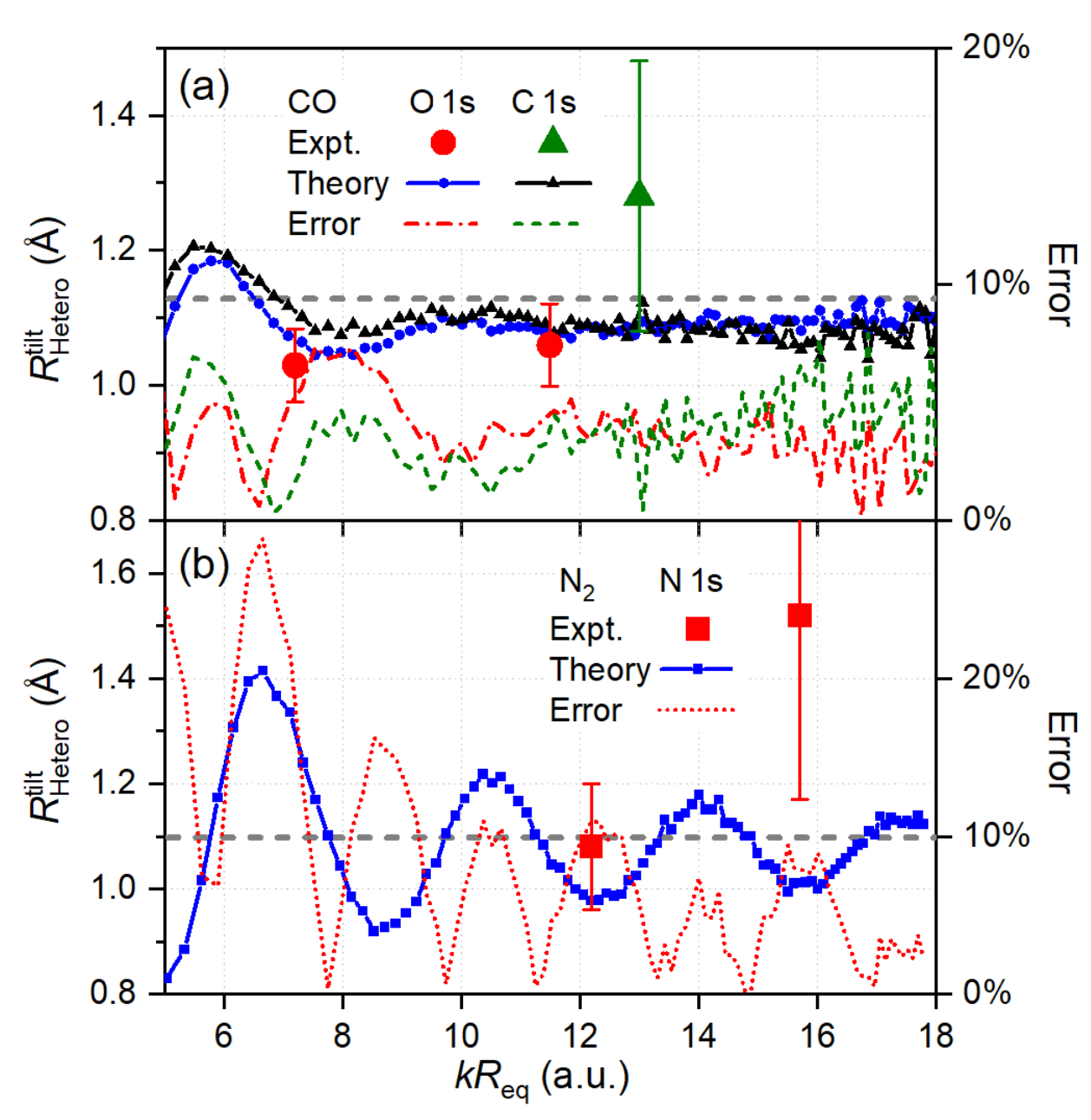}
\caption{Bond lengths (refer to the left-handed vertical axis), estimated from the tilt angles of the \textit{ab initio} computed   (solid lines with small symbols) and experimentally measured (large symbols with error bars) CP-MFPADs via the derived  relation~(\ref{eqn:tilt_hetero}) as functions of  $kR_\mathrm{eq}$, and the respective relative errors (broken lines, refer to the right-handed vertical axis). The equilibrium bond lengths  $R^{\rm CO}_{\rm eq}=1.1283$ and $R^{\rm N_{2}}_{\rm eq}=1.0977$~{\AA} are indicated by the  horizontal dashed lines. }
\label{fig:detR}
\end{figure}

Equation~(\ref{eqn:tilt_hetero}) connects directly the bond length of a molecule and the tilt angle $|\theta_{\rm tilt}|$ of the forward-scattering peak of the CP-MFPAD, and thus can be thought of as a simple, universal bond-length ruler. As pointed out in an insightful argument by Daimon and coworkers  \cite{Daimon1993}, the physics underlying Eq.~(\ref{eqn:tilt_hetero}) can be understood very intuitively. Upon birth the  photoelectron carries one unit of angular momentum which it inherited from the photon. Classically an electron moving with a momentum $\vec{k}$ along a straight line thus must follow a path of impact parameter $\vec{b}$ with respect to the nucleus such that $\vec{k}\cdot \vec{b}=1 \hbar$. This straight line defines the tilted symmetry axis of the photoemission. For $R>>b$ this results in $R=1/k \sin(\theta)$ and a symmetry axis of the problem which is tilted  by $\theta$ with respect to the molecular axis.  Figure \ref{fig:detR} depicts the bond length $R_{\rm Hetero}^{\rm tilt}$ estimated from the computed and measured CP-MFPADs using Eq.~(\ref{eqn:tilt_hetero}). The resulting values are plotted as functions of $kR_{\rm eq}$, where $R_{\rm eq}$=1.1283 and 1.0977 {\AA} for CO and N$_2$, respectively. Note that Eq.~(\ref{eqn:tilt_hetero}) is derived for hetero-nuclear diatomic molecules, and its application to the homo-nuclear N$_2$ molecule is a rather crude approximation, as we will discuss below. From Fig.~\ref{fig:detR}(a), it is evident that the bond lengths, estimated from both the {\it ab-initio}-modeled and measured CP-MFPADs, agree well with the equilibrium bond length of the neutral ground state of CO (the horizontal dashed line), with a deviation of less than 10\%. On the other hand, the bond lengths of N$_2$ in Fig.~\ref{fig:detR}(b), estimated from the {\it ab initio} CP-MFPADs, oscillate as a function of $kR_{\rm eq}$ around $R_{\rm eq}$. Those oscillations are damped and drop below 10\% for sufficiently high photoelectron energies. The bond length of N$_2$ estimated from the experimental CP-MFPADs is, however, in good agreement with the expected value of $R_{\rm eq}$ within the experimental uncertainties.

Please note that especially for high electron energies $\theta_{\rm tilt}$ becomes small and, accordingly, its extraction from the experimental data becomes very challenging. Small systematic errors can prevent the determination of the bond length in such cases. It turned out that performing a set of two measurements using both light helicities as references in the calibration can circumvent these problems to some extent. While in Fig. 1 all six measured MFPADs agree well with the computed ones, we were not in all cases able to extract the offset angle with the required precision. For N$_2$ at 780~eV, we recorded only one dataset with one helicity, its corresponding data point has a considerably larger error bar. The two measurements performed at the CO C 1s edge reflect a small systematic error due to overlapping flight times of the two ionic fragments, which directly (and almost solely) affected the part of the CP-MFPADs from which $\theta_{\rm tilt}$ is extracted. We included the lower-energy data point, with an estimate of the error which is caused by the overlapping flight times. Such experimental difficulties should be considered in future measurements employing the $\theta_{\rm tilt}$ approach.

The large oscillation in $R_{\rm Hetero}^{\rm tilt} $ in Fig.~\ref{fig:detR}(b) can be interpreted as follows. In the experiment, one is unable to distinguish  photoelectrons originating from the N 1s gerade and ungerade molecular orbitals of N$_2$, which are separated by $\sim 105$~meV~\cite{Ehara2006} (or from the degenerate 1s orbitals of the left and right nitrogen atoms). The resultant CP-MFPADs are therefore the sum of the two contributions:
\begin{eqnarray}
&\bar{I}\left(\bf{k},\hat{\varepsilon}_{\sigma}\right)& \propto
I_{g}\left(\bf{k},\hat{\varepsilon}_{\sigma}\right)
+
I_{u}\left(\bf{k},\hat{\varepsilon}_{\sigma}\right)
\nonumber\\
&&=
I^{A}\left(\bf{k},\hat{\varepsilon}_{\sigma}\right)
+
I^{A}\left(-\bf{k},\hat{\varepsilon}_{\sigma}\right).
\label{eqn:Igu}
\end{eqnarray}
Here, we used the fact that $f^A(k,\theta)=f^B(k,\theta+\pi)$ for homo-nuclear diatomic molecules.
One can see that Eq.~(\ref{eqn:Igu}) has a $C_2$ point symmetry, as is evident from Fig.~\ref{fig:tilt}(e).
The fringe function of Eq.~(\ref{eqn:Igu}) can thus be reduced to a superposition of fringe functions of two emitting nitrogen atoms
\begin{eqnarray}
H(\theta)
&\propto&
G(\theta)+G(\theta+\pi),
\end{eqnarray}
with $G(\theta)$ given by Eq.~(\ref{eqn:fringe}).
We see that it has also $C_2$ symmetry, namely $H(\theta)=H(\theta+\pi)$.
Because of the superposition of $G(\theta)$ and $G(\theta+\pi)$, the forward-tilted peak of $G(\theta)$ overlaps with the backward-scattering peak of $G(\theta+\pi)$. Since the backward-scattering exhibits EXAFS-type oscillations, the tilt angle of homo-nuclear diatomic molecules oscillates with $k$. These oscillations are damped with the energy as $\sim k^{-2}$, which corresponds to the asymptotic behavior of the backward-scattering amplitude~\cite{Schiff1968}.

In summary, we have investigated experimentally and theoretically the CP-MFPADs of core-level photoelectrons emitted from CO and N$_2$ molecules. We showed that (at high photoelectron energies) the tilt angle of the forward-scattering peak appearing in the CP-MFPAD of hetero-nuclear diatomic molecules can be directly connected to the bond length by a simple, universal equation. We applied the derived bond-length ruler formula to several examples of measured and computed CP-MFPADs. The extracted bond lengths agree well with the known bond lengths of CO and N$_2$ in their neutral ground states.  The present study illustrates that employing high-energy circularly polarized light, which is available not only at most synchrotron-radiation facilities but also at XFEL facilities, has many advantages for the extraction of the molecular structure. It allows to extract molecular bond lengths not only on a qualitative, but on a quantitative level. We assume that our findings hold not only for the static case, but also for time-resolved MFPAD measurements aiming at catching the motion of individual atoms of molecules undergoing photoreactions, as e.g. the stretching of chemical bonds. \\

\begin{acknowledgements}
The present work was funded in part by the Deutsche Forschungsgemeinschaft (DFG) Project No. 328961117-SFB 1319 ELCH (Extreme light for sensing and driving molecular chirality, subprojects B1 and C1). F.O., K.H., and K.U. acknowledge Cooperative Research Program of "Network Joint Research Center for  Materials and Devices". K.H. also acknowledges funding by JSPS KAKENHI under Grant No. 18K05027 and 19KK0139, while K.U. also acknowledges the X-ray Free Electron Laser Utilization Research Project and the X-ray Free Electron Laser Priority Strategy Program of the Ministry of Education, Culture, Sports, Science, and Technology of Japan (MEXT) and the IMRAM program of Tohoku University. D.Y. acknowledges JSPS KAKENHI Grant Number JP19J12870 and a Grant-in-Aid of Tohoku University Institute for Promoting  Graduate Degree Programs Division  for Interdisciplinary Advanced Research and Education. K.Y. is grateful to the financial support from JSPS KAKENHI Grant Number 19H05628. We acknowledge the SOLEIL synchrotron facility for provision of synchrotron radiation and we would like to thank N. Jaouen and his team for assistance in using beamline SEXTANTS under proposal 20170599.

M.S.S., T.J., R.D., P.V.D., and K.U. conceived the present work. I.V.-P., F.O., and A.M. contributed to it equally. The experiment was prepared and carried out by I.V.-P., J.R., N.M., S.U., G.N., N.A., D.Y., R.G., M.N.P., M.S., V.T.D., M.K., C.J., M.W., J.B.W., K.U., R.D., M.S.S., and T.J.. Experimental data analysis was performed by I.V.-P., K.F., and T.J.. {\it Ab initio} calculations were performed by A.M. and P.V.D.. Analytical formulas were derived and employed for the analysis by F.O., Y.T., and K.H.. F.O., Y.T., K.H., K.Y., K.U., P.V.D., R.D.,  M.S.S., and T.J. wrote the paper. All authors discussed the results and commented on the manuscript.
\end{acknowledgements}

\bibliographystyle{apsrev4-2}

\end{document}